\documentclass [12pt,a4paper]{article}
\usepackage {graphicx}
\begin{document}
\author{}
\title{
\large\bf Systematic Field-Theory for the Hard-Core One-Component
Plasma\\ \phantom {empty line} \normalsize Nikolai V.
Brilliantov$^{1,2}$, Vitali V. Malinin$^{2}$ and Roland R.
Netz$^{1}$\\ \small\it $^1$ Max-Planck-Institute of Colloids and
Interfaces, 14424 Potsdam, Germany\\ \small\it $^2$ Moscow State
University, Physics Department, Moscow 119899, Russia\\}
\date{}
\maketitle

\begin{abstract}

An accurate and systematic equation of state for the hard-core
one-component plasma (HCOCP) is obtained. The result is based on
the Hubbard-Schofield transformation which yields the
field-theoretical Hamiltonian, with coefficients expressed in
terms of equilibrium correlation functions of the reference
hard-core fluid. Explicit calculations were performed using the
Gaussian approximation for the effective Hamiltonian and known
thermodynamic and structural properties of the reference hard-core
fluid. For small values of the plasma parameter $\Gamma$ and
packing fraction the Debye-H\"uckel result is   recovered, while
for $\Gamma\gg 1$, the excess free energy $F_{ex}$ and internal
$U_{ex}$ energy depend linearly on   $\Gamma$. The obtained
expression for $U_{ex}$ is in a good agreement with the available
Monte Carlo data for the HCOCP. We also analyse the validity of
the widely used approximation, which represents the free energy as
a sum of the hard-core and electrostatic part.
\end{abstract}

\par\bigskip
Preprint accepted to Journal of Physics D: Applied Physics.
\par\bigskip

\section{Introduction}

The one component plasma (OCP) is one of the basic models in the
field of charged systems. The OCP model is formulated as a system
of point particles, interacting via the Coulomb potential, which
move in a uniform neutralizing background. It has found important
physical applications in a variety of fields ranging from
terrestrial physics, through important technological applications
to cosmology \cite{on1,on2,on3,on4}. As a reference model it is
used in many areas of soft condensed matter, such as colloidal and
polyelectrolyte solutions, e.g.
\cite{StevRob1990,Lowen1994,PenJonNord1993,BKK1998}, etc. All
thermodynamic properties of the OCP depend only on the
dimensionless plasma parameter $\Gamma=l_B/a_c$, where
$l_B=e^2/k_B T$ is the Bjerrum length ($e$ is the charge of the
particles, $k_B$ is the Boltzmann's constant, $T$ is the
temperature) and $a_c=(3/4\pi\rho)^{1/3}$ is the ion-sphere radius
with $\rho=N/\Omega$ being the concentration of particles ($N$ is
the number of particles, $\Omega$ is the volume of the system).
Using a field-theoretical approach, a  fairly accurate and simple
expression for the equation of state of  the OCP has been obtained
within the Gaussian approximation for the effective Hamiltonian
\cite{brocp}; contrary to previous calculations, e.g.
\cite{nord,SWS}, this gives a correct behavior for the
thermodynamic functions in the full range of $\Gamma$ (see
\cite{brocp} and references therein) and does not have fitting
parameters as e.g. in \cite{SWS,BraHanJo1979}. In
\cite{MoreiraNetz2000} field-theoretical calculations for  the
equation of state, going beyond the Gaussian approximation, have
been performed, showing that corrections to the Gaussian theory
are rather small.

A closely related model  -- the hard-core one component plasma
(HCOCP) incorporating a hard core repulsion between ions -- gives
a more satisfactory description of the short-range electrostatic
correlations. The importance of this model follows also from the
fact that it belongs to the class of the  so-called primitive
models used to describe molten salts \cite{tw7}, electrolytes
\cite{tw8}, liquid metals \cite{on40,on41,on42} and charged
colloidal solutions \cite{tw9,tw10,tw11}. The mean spherical
approximation (MSA) \cite{on30} was applied to account for both
the hard-core and electrostatic interactions in this system
\cite{pen31,pen32} . Although it has an analytical solution, it
exhibits sometimes unphysical negative contact values for the pair
correlation function, which has to be remedied by a rescaling
procedure\cite{on52,on53,on54}. A mixed Percus-Yevick/hypernetted
chain integral equation also has been used \cite{on33} and it was
observed that it fits better the simulation results \cite{on34}
than the MSA and cluster expansion\cite{on35,on36}. The former
however does not have an analytical formulation. An analytical
equation of state for the HCOCP has been proposed in \cite{hs1},
as a simple generalization of the hole-corrected Debye-H\"uckel
theory \cite{nord}, which in addition to the correlation hole
around charged particles takes into account the hard-core
repulsion. This was used afterwards to develop a generalized van
der Waals theory \cite{hs2}. The basic physical idea exploited in
this theory is that due to the strong electrostatic repulsion a
``hole''  appears around a charged particle from which all other
particles are expelled. Outside the hole the electrostatic
interactions are not very strong and may be described on the
Debye-H\"uckel level; the size of the hole is found
self-consistently. Although being physically appealing, this
theory does not give a satisfactory description for large values
of the plasma parameter $\Gamma$  for the OCP  \cite{nord}, and is
also not accurate for the HCOCP for large packing fractions $\eta$
(for $\eta=0.4$ the deviations from MC data for the internal
energy reach $24\%$ \cite{hs1}). Recently {\it an exact}
low-density expansion for the free energy of the HCOCP has been
obtained \cite{NetzOrland2000}. This however can not be applied in
the case of strong electrostatic interaction, i.e. for large
$\Gamma$.

In the present study we develop a theory which allows to derive a
fairly accurate and simple equation of state for the HCOCP in the
whole range of $\Gamma$ from the Debye-H\"uckel limit $\Gamma \ll
1 $ up to the limit of strong coupling $\Gamma\gg 1$. It
reproduces within 1-3\% accuracy the available Monte Carlo (MC)
data for $0.2<\Gamma<70$. Larger deviations occur in the region
where the MC data are not very accurate (see the discussion
below). As in the case of the OCP \cite{brocp} we use the
Hubbard-Schofield transformation which yields a field-theoretical
Hamiltonian for the HCOCP with coefficients expressed in terms of
equilibrium correlation functions of a reference hard-core fluid.
Although we perform the explicit calculations within the Gaussian
approximation for the effective Hamiltonian, one can go beyond the
Gaussian approximation using the standard perturbation technique,
provided the structural properties of the reference hard-core
fluid are known. Since the obtained equation of state is in good
agreement with available MC data we use it to address the problem
of additivity of the hard-core and electrostatic contributions to
the excess free energy. This seems to be important since it is
usually assumed that these contributions are additive (e.g.
\cite{levin98}) and the range of validity of such approximation
has not been studied yet.

The rest of the paper is organized as follows: in  Sec.2 we
briefly sketch the Hubbard-Schofield approach which yields  the
field-theoretical Hamiltonian for the HCOCP. In Sec.3 we derive
the equation of state within the Gaussian approximation for the
effective Hamiltonian, and compare the theoretical findings with
available Monte Carlo data. In that section we also analyse the
accuracy of the widely used approximation in which the excess free
energy is written as a sum of the hard-core and electrostatic
part. In the Conclusion we summarize our findings.

\section{Field-Theoretical Hamiltonian for HCOCP}

We start from the HCOCP Hamiltonian with omitted ideal part which
may be written as follows ($\beta^{-1}=k_BT$):
\begin{equation}
\label{1}
H=\frac{1}{2}\beta^{-1}\displaystyle\mathop{{\sum}'}_{\bf k}\nu_k
(\rho_{\bf k}\rho_{-\bf k}-\rho)+H_{hc}
\end{equation}
where the first term in the right-hand side of (\ref{1}) refers to
the Coulomb interactions, written in terms of collective density
variables,
\begin{equation}
\label{2} \rho_{\bf k}=\frac{1}{\sqrt{\Omega}} \sum_{j=1}^N
e^{-i{\bf kr}_j}
\end{equation}
where ${\bf r}_j$ denotes coordinate of j-th particle,
\begin{equation}
\label{nukdef} \nu_k=4\pi l_B/k^2
\end{equation}
is the Fourier-transformed Coulomb potential and $H_{hc}$
describes the hard-sphere interaction. Summation in (\ref{1}) is
to be performed over the wave vectors ${\bf k}= \{k_xk_yk_z\}$
with $k_i=2\pi l_i/L$ $(i=x,y,z)$, where $l_i$ are integers,
$L^3=\Omega$, and the prime over the sum denotes that the term
with ${\bf k}=0$ is excluded \cite{explain1}.

\subsection{Hubbard-Schofield transformation}

The configurational integral may be written in terms of the
configurational integral of the reference (hard-sphere liquid)
system $Q_R$ \cite{br30,br31,br32} as:
\begin{equation}
\label{3} Q=\left< \exp\left\{-\frac{1}{2}\mathop{{\sum}'}_{\bf
k}\nu_k (\rho_{\bf k}\rho_{-\bf k}-\rho)\right\}\right>_R Q_R
\end{equation}
where $\left<(\dots)\right>_R=Q^{-1}_R\int d{\bf r}^N(\dots)$
denotes the averaging over the reference system. In accordance
with the Hubbard-Schofield scheme \cite{br30} we use the identity:
$$ \exp\left (\frac{1}{2}a^2x^2\right)=\frac{1}{\sqrt{2\pi a^2}}
\int\limits^{+\infty}_{-\infty}\exp\left
(-\frac{1}{2}y^2/a^2+ixy\right)dy $$
and arrive after some algebra at:
\begin{equation}
\label{Q_start} Q=Q_R\int \mathop{{\prod}'}_{\bf k}c_{\bf
k}d\varphi_{\bf k} \exp\left\{-\frac{1}{2}\mathop{{\sum}'}_{\bf
k}\nu_k^{-1} \varphi_{\bf k}\varphi_{-\bf k}\right\}
\left<\exp\left\{i\mathop{{\sum}'}_{\bf k}\rho_{\bf
k}\varphi_{-\bf k}\right\} \right>_R
\end{equation}
where $c_{\bf k}=(2\pi \nu_k)^{-1/2}\exp\{\nu_k \rho/2\}$, and
where the integration is to be performed under restriction,
$\varphi_{-\bf k}=~\varphi^{*}_{\bf k}$ ($\varphi^{*}_{\bf k}$ is
the complex conjugate of $\varphi_{\bf k}$) \cite{explain2}.
Applying the cumulant theorem \cite{kubo} to the factor
$\left<\exp\left\{i\mathop{{\sum}'}_{\bf k} \rho_{\bf
k}\varphi_{-\bf k}\right\}\right>_R$ one obtains:
\par\bigskip
\centerline{$Q=Q_R\displaystyle \int \mathop{{\prod}'}_{\bf
k}c_{\bf k} d\varphi_{\bf k}e^{-{\cal H}}$, \phantom{*} with
\phantom{**********}}
\begin{equation}
\label{5} {\cal
H}=\sum^{\infty}_{n=2}\Omega^{1-\frac{n}{2}}\mathop{{\sum}'}_{{\bf
k}_1, \dots {\bf k}_n} u_n({\bf k}_1,\dots {\bf k}_n)\varphi_{{\bf
k}_1}\dots \varphi_{{\bf k}_n}
\end{equation}
$$ u_2({\bf k}_1,{\bf k}_2)=\frac{1}{2}\delta_{{\bf k}_1+{\bf
k}_2,0} \left\{\frac{1}{\nu_k}+\left<\rho_{{\bf k}_1}\rho_{-{\bf
k}_1}\right>_{cR} \right\} $$
$$ \phantom{***} u_n({\bf k}_1,\dots {\bf k}_n)=-i^ n
\frac{\Omega^{\frac{n}{2}-1}}{n!} \left<\rho_{{\bf
k}_1}\dots\rho_{{\bf k}_n}\right>_{cR} \phantom{**}n>2 $$
here $\left<\dots\right>_{cR}$ denotes {\it cumulant} average
\cite{kubo} for the reference hard sphere fluid system. Note that
(\ref{5}) gives the field-theoretical expression for the partition
function with, $\varphi_{\bf k}$ being the Fourier components of
the scalar field $\varphi(\vec r)$ \cite{Ma}. The coefficients of
the effective Hamiltonian (\ref{5}) are expressed in terms of the
correlation functions of the reference hard-core fluid. Using
definitions of correlation functions of fluids \cite{Gubbins} and
definitions of the cumulant averages \cite{kubo}, one can {\it
directly} evaluate $\left<\rho_{{\bf k}_1}\dots\rho_{{\bf
k}_n}\right>_{cR}$ (and thus the coefficients $u_n({\bf k}_1,\dots
{\bf k}_n)$). It is straightforward to show that $\left<\rho_{{\bf
k}_1}\dots\rho_{{\bf k}_n}\right>_{cR}$ may be expressed in terms
of the Fourier transforms of the correlation functions
$h_2,h_3,\dots,h_n$ of the reference system, defined as
\cite{br32}
\begin{equation}
\label{cor_h2} h_2({\bf r}_1,{\bf r}_2)\equiv g_2({\bf r}_1,{\bf
r}_2)-1
\end{equation}
\begin{equation}
\label{cor_h3} h_3({\bf r}_1,{\bf r}_2,{\bf r}_3) \equiv g_3({\bf
r}_1,{\bf r}_2,{\bf r}_3)-g_2({\bf r}_1,{\bf r}_2)- g_2({\bf
r}_1,{\bf r}_3)-g_2({\bf r}_2,{\bf r}_3)+2\,, \mbox{etc.,}
\end{equation}
where $g_l({\bf r}_1,..,{\bf r}_l)$ -- are $l$-particle
correlation functions \cite{Gubbins}. In particular the
second-order cumulant $\left<\rho_{\bf k}\rho_{-{\bf
k}}\right>_{cR}$ may be written as:
\begin{equation}
\label{cum_2} \left<\rho_{\bf k}\rho_{-{\bf k}}\right>_{cR} =\rho
\left [1+\rho \tilde h_2({\bf k})\right ]\,,
\end{equation}
and the Fourier transform of the function $\tilde h_2({\bf k})$ is
related to the direct correlation function $\tilde c_2({\bf k})$
\begin{equation}
\label{h2_k} \tilde h_2({\bf k})=\tilde c_2({\bf k})/[1-\rho\tilde
c_2({\bf k})]\,,
\end{equation}
for which an explicit analytical expression is known \cite{explain4}.

\subsection{Gaussian approximation for the effective Hamiltonian}

Now we concentrate on the Gaussian part of the effective
Hamiltonian, i.e. we skip all the terms with a power of the field
larger than two. The accuracy of this approximation has been
critically examined in Ref.\cite{MoreiraNetz2000} for the OCP
without hard-core interactions by perturbatively calculating
higher-order terms. It was found that higher order terms
contribute very little to the free energy and that the Gaussian
approximation is  in fact excellent over the whole range of
coupling parameters. One can therefore assume, and this is indeed
borne out by our comparison with Monte-Carlo data below, that the
Gaussian approximation should also be quite good for the present
case. Using (\ref{h2_k}) and (\ref{cum_2}) we write for this case:
\begin{equation}
\label{Q_gauss} Q=\displaystyle Q_R\int \mathop{{\prod}'}_{\bf
k}d\varphi_{\bf k} \frac{\exp (\nu_k\rho/2)}{\sqrt{2\pi \nu_k}}
\exp\left\{-\frac{1}{2}\mathop{{\sum}'}_{\bf k} \varphi_{\bf
k}\varphi_{-\bf k} \left (\frac{1}{\nu_k}+ \frac{\rho}{1-\rho
\tilde c_2({\bf k})}\right )\right\}
\end{equation}
Performing (Gaussian) integration over $\varphi_{\bf k}$ we arrive
after some algebra at \cite{explain2}:
\begin{equation}
\label{Q_fin} Q=\displaystyle
Q_R\mathop{{\prod}'}_{k_z>0}\frac{\exp (\nu_k\rho)} {\nu_k}\left
(\frac{1}{\nu_k}+ \frac{\rho}{1-\rho \tilde c_2({\bf k})}\right
)^{-1}
\end{equation}
Taking the logarithm of the configuration integral one obtains the
free energy, and since  the analytical expression for $\tilde
c_2({\bf k})$ is available \cite{explain4}, no additional
approximation is, in principle, required. This leads, however, to
an expression which is to be evaluated numerically. We note that
owing to the long-range nature of the Coulombic interactions, the
main contribution to the free energy comes from the long-wave
modes of the density fluctuations, which correspond to small $k$.
Therefore only the small-$k$ behavior of the direct correlation
function is important. This suggests to approximate $\tilde
c_2({\bf k})$ by a truncated expansion
\begin{equation}
\label{c2_k} \tilde c_2({\bf k}) \simeq  \tilde c_2(0)-\tilde
c_2(0)''k^2+\dots    \, ,
\end{equation}
which correctly behaves at small $k$. As we show in what follows,
only wave-vectors with $k < k_0$ contribute to the configuration
integral, thus  we apply the approximation (\ref{c2_k}) for the
interval $0<k<k_0$. By numerical evaluation of the free energy
using the full expression for $\tilde c_2({\bf k})$, we convinced
ourselves that deviations of the quadratic form from the actual
$\tilde c_2({\bf k})$ at larger $k$ do not noticeably affect the
results\cite{comment}. We therefore do not require $k$ and  $k_0$
to be small since the particular behavior of $\tilde c_2({\bf k})$
for large $k$ is not important. On the other hand, the quadratic
approximation allows to obtain an analytical equation of state for
the HCOCP that  reproduces fairly well the available Monte Carlo
data.  With (\ref{c2_k}) one can write  for the configurational
integral:
\begin{equation}
\label{Q_small} Q=\displaystyle
Q_R\mathop{{\prod}'}_{k_z>0}\frac{\exp (\nu_k\rho)} {\nu_k}\left
(\frac{1}{\nu_k}+ \frac{\rho}{1-\rho \tilde c_2(0)}-
k^2\frac{\rho^2\tilde c_2(0)''}{(1-\rho \tilde c_2(0))^2}\right
)^{-1}
\end{equation}
The most accurate estimate for $\tilde c_2(0)$ may be found, using
$\tilde c_2(0)=\tilde h_2(0)/(1+\rho \tilde h_2(0))$ from
(\ref{h2_k}) and the relation for the isothermal compressibility
$\chi_R^{-1}=\rho(\partial P_R/\partial \rho)_\beta$
\cite{Gubbins}, where $P_R$ is the pressure of the reference
system \cite{c0comm}:
\begin{equation}
\label{Z_0} 1+\rho\tilde h_2(0)=\rho k_BT\chi_R\equiv Z_0
\end{equation}
The value of $Z_0$ follows from the fairly accurate
Carnahan-Starling free energy of hard-sphere fluid \cite{CarnHS}:
\begin{equation}
\label{Fhc} \frac{\beta
F_{hc}}{N}=\frac{4\eta-3\eta^2}{(1-\eta)^2}\,,
\end{equation}
where $\eta = \frac16 \pi \rho d^3$ is  the packing fraction and
$d$ is the diameter of the spheres; this yields $Z_0$ as a second
derivative of $F_{hc}$ with respect to density:
\begin{equation}
\label{Z_eq} Z_0=(1-\eta)^4(1+4\eta+4\eta^2-4\eta^3+\eta^4)^{-1}
\end{equation}
Using the Wertheim-Thiele solution for the direct correlation
function $c_2(r)$ \cite{explain4} and definition of $c_2(0)''$
from Eq.(\ref{c2_k}) one obtains:
\begin{equation}
\label{c_2''} \tilde {c}_2(0)''=\frac{1}{2}\int r^2c_2(r)d{\bf r}=
-(\pi d^5/120)(16-11\eta+4\eta^2)(1-\eta)^{-4}
\end{equation}
With $\tilde {c}''(0)$ from (\ref{c_2''}) and $\tilde
{c}_2(0)=(Z_0-1)/(\rho Z_0)$ from (\ref{h2_k}) and (\ref{Z_0}) we
arrive at the following expression for the configurational
integral
\begin{equation}
\label{Q_end} Q=\displaystyle Q_R\mathop{{\prod}'}_{k_z>0} \exp
(\rho\nu_k)\left (\rho\nu_k Z_0+\Theta\right )^{-1}
\end{equation}
where $\Theta=1-4\pi l_B(\rho Z_0)^2\tilde c_2''(0)$.

\section{Equation of state for the HCOCP}

\subsection{Equation of state and its accuracy}

Now we show that within the Gaussian approximation to the
effective Hamiltonian, one obtains rather accurate equation of
state for HCOCP provided that an appropriate value of the
``ultraviolet cutoff'' in the ${\bf k}$-space is employed. From
(\ref{Q_end}) we find for the excess free energy of the HCOCP:
\begin{equation}
\label{Fex} -\beta F_{ex}=\ln(Q)=-\beta
F_{hc}+\frac{1}{2}\mathop{{\sum}'}_{\bf k}
[\rho\nu_k-\ln(\rho\nu_k Z_0+\Theta)]
\end{equation}
Now we argue that the summation in (\ref{Fex}) should be carried
out over a {\it finite} number of the wave-vectors ${\bf k}$. In
this we follow the Debye theory of the specific heat of solids
(e.g. \cite{br42}). Namely, we assume that the total number of
degrees of freedom in the system, $3N$, should be equal to the
total number of {\it physically different} modes with the
wave-vectors ${\bf k}$ within the spherical shell of radius $k_0$
in the ${\bf k}$-space. The number of modes is twice the number of
the wave-vectors, since for each ${\bf k}$ one has a sine and
cosine mode (the amplitude of the ${\bf k}$-th mode is a complex
number) \cite{explain5}. Thus we obtain:
\begin{equation}
\label{k0_def} 2\frac{\Omega}{8\pi^3}4\pi\int^{k_0}_0 k^2dk=3N
\end{equation}
where the factor $\Omega/8\pi^3$ appears when the integration in
${\bf k}$-space is used instead of summation. From (\ref{k0_def})
follows that
$$ k_0=(9\rho\pi^2)^{1/3} $$
A similar Debye-like scheme to find the cutoff $k_0$ was first
proposed for plasma in \cite{br44}, where a somewhat different
value of the cutoff wave-vector was reported. Using $k_0$ as
obtained above we write:
\begin{equation}
\begin{array}{l}
\label{13} \displaystyle\frac{\beta F_{ex}}{N}=\frac{\beta
F_{hc}}{N}+\frac{1}{2}
\frac{\Omega}{8\pi^3}\frac{4\pi}{N}\int\nolimits^{k_0}_{0}k^2dk
\left[\ln\left(\rho\nu_k Z_0+\Theta\right)-\rho\nu_k\right]\\
\phantom{*}\\ \displaystyle\qquad=\frac{\beta
F_{hc}}{N}+\frac{9}{4}\int\nolimits^1_0x^2dx
\left[\ln\left(\Theta+\frac{b\Gamma
Z_0}{x^2}\right)-\frac{b\Gamma}{x^2}\right]\,,
\end{array}
\end{equation}
where $x=k/k_0$, so that $\rho\nu_k=4\pi l_B
\rho/(k_0x)^2=b\Gamma/x^2$ (see (\ref{nukdef}) for the definition
of $\nu_k$), and where we define the constant
$b=\frac{2}{3}\left(\frac{2}{\pi^2}\right)^{1/3}$. The last
integral is easily evaluated  and yields for the excess free
energy
\begin{equation}
\label{Fex_fin}
\frac{F_{ex}}{k_BTN}=\frac{4\eta-3\eta^2}{(1-\eta)^2}+\frac{3}{4}
\left[\ln\left(\Theta+b\Gamma
Z_0\right)-b\Gamma\left(3-\frac{2Z_0}{\Theta}
\right)\right]-\frac{3}{2}\left(\frac{b\Gamma
Z_0}{\Theta}\right)^{\frac{3}{2}}
\arctan\left(\sqrt{\frac{\Theta}{b\Gamma Z_0}}\phantom{|}\right)
\end{equation}
with
\begin{equation}
\label{15} \Theta=1+\frac{6}{5}\frac{e^2}{d k_B T}
\frac{\eta^2(1-\eta)^4(16-11\eta+4\eta^2)}{(1+4\eta+4\eta^2-4\eta^3+\eta^4)^2}
\end{equation}
and for the excess internal energy,
$
U_{ex}=-T^2\partial(F_{ex}/T)/\partial T
$
of the HCOCP
\begin{equation} \label{U_fin}
\frac{U_{ex}}{k_BTN}=\frac{9}{4}\left [\frac{b\Gamma
Z_0}{\Theta}-b\Gamma- \left (\frac{b\Gamma
Z_0}{\Theta}\right)^{\frac{3}{2}}\arctan
\left(\sqrt{\frac{\Theta}{b\Gamma Z_0}}\,
\right)\right]\,.
\end{equation}
For $\eta=0$,  Eqs.(\ref{Fex_fin},\ref{U_fin}) recover the
corresponding result for the one component plasma \cite{brocp}.

As it follows from (\ref{Fex_fin}) and (\ref{U_fin}), for
$\Gamma\to 0$ and $\eta\to 0$ the Debye-Huckel behavior is
obtained. In the opposite limit $\Gamma\gg 1$ and for any packing
fraction $\eta$ Eqs.(\ref{Fex_fin}) and (\ref{U_fin}) demonstrate
a linear behavior on $\Gamma$. The leading term for this case  is
$-A\Gamma$, where the constant $A$ reads:
$$
A=\frac{9}{4}b=\frac{3}{2}\left(\frac{2}{\pi^2}\right)^{1/3}=0.881\dots
$$
This is fairly close to the constant $A=0.899\dots$ of the fits
for the OCP (see e.g.\cite{br4,br5}).

For arbitrary values of $\eta$ and $\Gamma$ we compare our
analytical expression for the excess internal energy (\ref{U_fin})
with the available MC data for the HCOCP (Fig.1). Fig.2 shows the
relative error of the analytical expression (\ref{U_fin}). As it
follows from Fig.1 and Fig.2 the equation of state is fairly
accurate in full range of plasma parameters for which MC data are
available. For $\Gamma>1$ the relative error does not exceed
$1-3\%$ for all values of packing fraction and plasma parameter.
For $\Gamma>10$ one observes the linear behavior and deviation
from the numerical data for this range of $\Gamma$ less than
$1\%$.

The maximal deviation of the analytical expression from the MC
data occurs for $0.2<\Gamma<1$. In this range of the plasma
parameter the deviation is about $5\%$ with the maximal one of
$12\%$ at the smallest value $\Gamma=0.2$.  It should be noted
however, that such deviation occurs for $\Gamma$, where  the
method of MC loses its accuracy. Moreover, we expect that for very
small packing fraction, $\eta=0.001$, where the maximal deviation
is observed, the contribution to the internal energy due to the
hard-core interactions may not exceed $1\%$. Therefore for this
$\eta$ the difference between $U_{ex}$ of the HCOCP and $U_{ex}$
of the OCP is less then $1\%$. It has been already shown that the
analytical equation of state for the OCP (which follows from
(\ref{U_fin}) for $\eta=0$) has the accuracy of $8\%$
\cite{brocp}. Thus we expect that the accuracy of our equation of
state may not be worse than $9\%$ for these values of $\eta$; the
observed deviation of $12\%$ seems to be the manifestation of the
low accuracy of MC for this range of parameters.

\subsection{Additivity of the hard-core and electrostatic
components of the free energy.}

With the equation of state obtained we can analyse the accuracy of
the widely used approximation for the excess free energy, which
represents this as a sum of the hard-core and electrostatic
component, $F_{hc}+F_{ex,OCP}$ (see e.g. \cite{levin98}). The
validity of this approximation may be checked by the direct
comparison of the excess free energy of the HCOCP $F_{ex,HCOCP}$
and the above sum. For $F_{ex,HCOCP}$ we use our equation of state
(\ref{Fex_fin}) and equation of state for OCP from \cite{brocp}
(this expression for $F_{ex,OCP}$ may be obtained from
Eq.(\ref{Fex_fin}) for $\eta=0$). For $F_{ex,hc}$ we use the
Carnahan-Starling Eq.(\ref{Fhc}). In Fig.3 we compare  the
''complete'' equation of state (solid lines) with the approximate,
based on the assumption of additivity of the hard-core and
electrostatic parts (dashed lines). As it follows from Fig.3 for
small $\Gamma$ the values of the free  energy from the
''complete'' and approximate equations of state may differ
significantly; they may even have different signs. The relative
difference becomes smaller with increasing $\Gamma$ (see inset to
Fig.3) and decreasing packing fraction. For $\Gamma>50$ the
approximation reaches the reasonable accuracy: the error does not
exceed $5\%$ for all $\eta$. For small packing fractions,
$\eta<0.1$ one has the same accuracy already for $\Gamma>10$.

\section{Results and Discussion}

A "first-principle" equation of state for the classical hard-core
one component plasma is obtained that has a correct Debye-Huckel
behavior at the limit of small plasma parameter $\Gamma$ and small
packing fractions ($\Gamma \to 0$, $\eta \to 0$). It demonstrates
a linear dependence on $\Gamma$ for any packing fraction if
$\Gamma \gg 1$. The obtained coefficient $0.881$ at the linear
leading term in this case is close to the corresponding
coefficient $0.899$ found for the one component plasma in the
Monte Carlo simulations. The simple analytical expression for the
excess internal energy reproduces the available MC data with an
accuracy of $1-3\%$ for the most range of $\Gamma$ and $\eta$. The
maximal deviation of $12\%$ is observed for small $\Gamma$ and
$\eta$, where the MC method is not very accurate, and we argue
that such deviation does not reflect the accuracy of our equation.

To derive the equation of state we apply  the Hubbard-Schofield
transformation to obtain the field theoretical Hamiltonian for the
HCOPC and use the Gaussian approximation. The Gaussian
approximation assumes that one can neglect all terms in the
effective Hamiltonian which contain the power of field higher than
two. As it was shown by the field theoretical calculations for the
case of the one component plasma (without the hard-core)
\cite{MoreiraNetz2000}, the corrections to the Gaussian theory are
rather small, and we expect that the same is true for the system
of interest. Physically, this implies that the higher-order
coefficients in the field theoretical Hamiltonian are smaller than
that at the second order of the field. In this case the
contribution to the field integral from the field-amplitude domain
where the higher-order terms dominate is small: The main
contribution comes from the domain where the Gaussian term
prevails. Generally, the Gaussian approximation fails near the
critical point, where the Gaussian term of the effective
Hamiltonian becomes small and even vanishes \cite{br32}. For the
HCOCP we do not expect any criticality and thus the Gaussian
approximation is expected to be valid.

To obtain the analytical expression for the equation of state we
also approximate the Fourier transform of the direct correlation
function of the reference hard-core system $\tilde c_2({\bf k})$
by its small-$k$ expansion. Although such approximation deviates
from the actual dependence of $\tilde c_2({\bf k})$ at larger $k$,
this does not eventually affect the equation of state: Owing to
the long-range nature of the Coulombic interactions, only the
small-$k$ behavior of $\tilde c_2({\bf k})$ is important; this is
correctly reproduced by the approximation.

We also analyse the validity of the widely used approximation,
where the free energy of the HCOCP is represented as a sum of the
hard-core and electrostatic component. We show that such
approximation is rather accurate for small packing fraction and
large plasma parameter. In the opposite case of large $\eta$ and
small $\Gamma$ the excess free energy may not be adequately
represented by a sum of a hard-core and electrostatic part.

Thus we conclude that in general case one has to use the complete
equation of state. The proposed one possesses a reasonable
accuracy for the whole range of parameters for the system of
interest.

\newpage

\begin{figure}[ht]
\centering
\includegraphics[scale=0.5,angle=270]{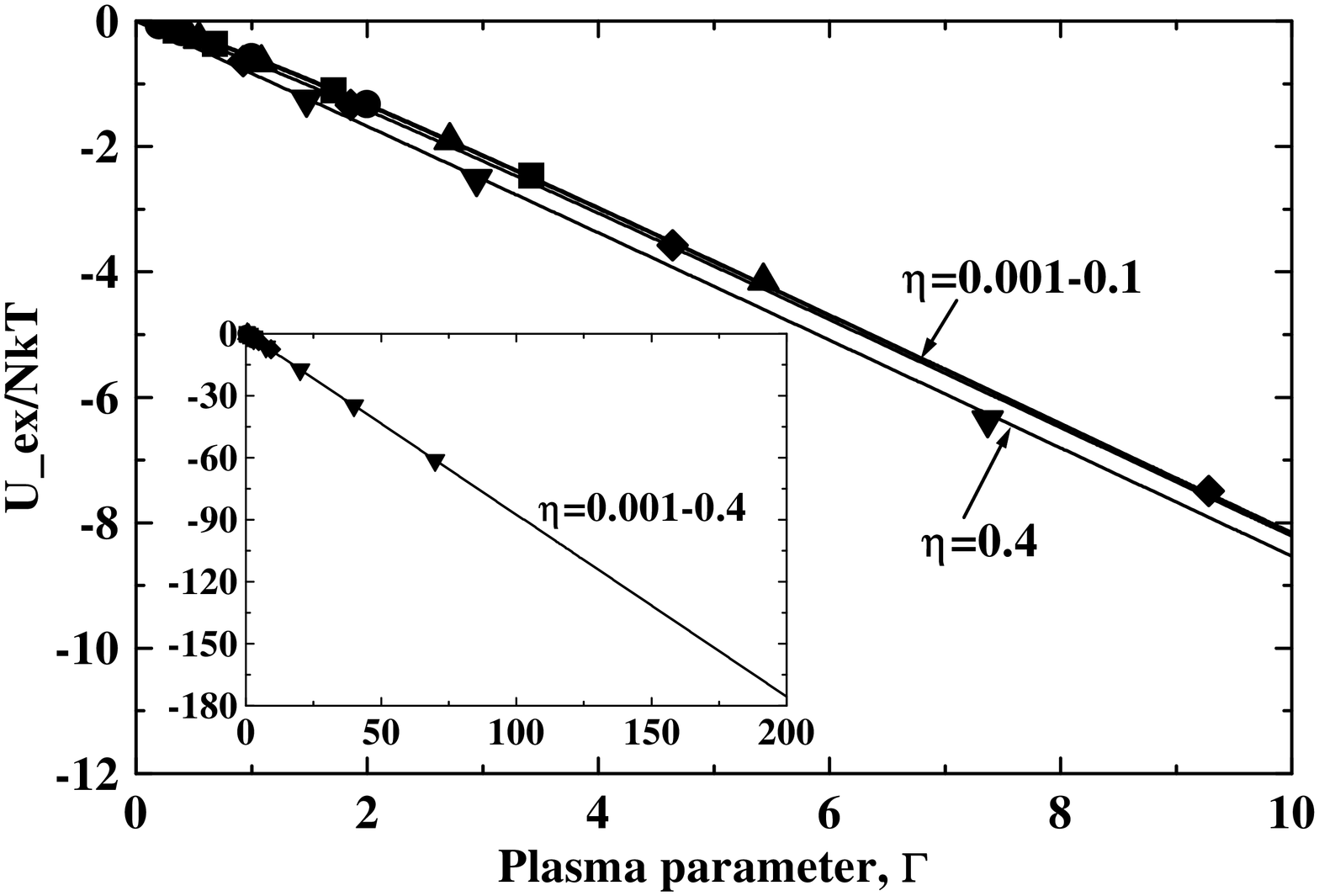}
\caption{Shows the dependence of the excess internal energy of the
HCOCP $U_{ex}/Nk_BT$ on the plasma parameter $\Gamma=l_B/a_c$
($l_B=e^2/k_B T$, $a_c=(3/4\pi\rho)^{1/3}$) for different values
of the packing fraction $\eta=(\pi/6) d^3\rho$ .The curves from
top to bottom correspond to $\eta=0.001$, $\eta=0.005$,
$\eta=0.020$, $\eta=0.100$ and $\eta=0.400$, respectively. Points
give the Monte-Carlo data \protect \cite{on34,hs1}: circles
correspond to $\eta=0.001$, squares to $\eta=0.005$, triangles to
$\eta=0.020$, diamonds to $\eta=0.100$ and down triangles to
$\eta=0.400$, respectively. In the inset the same dependence is
shown for larger range of $\Gamma$.} \label{tot_hcen}
\end{figure}

\newpage

\begin{figure}[ht]
\centering
\includegraphics[scale=0.5,angle=270]{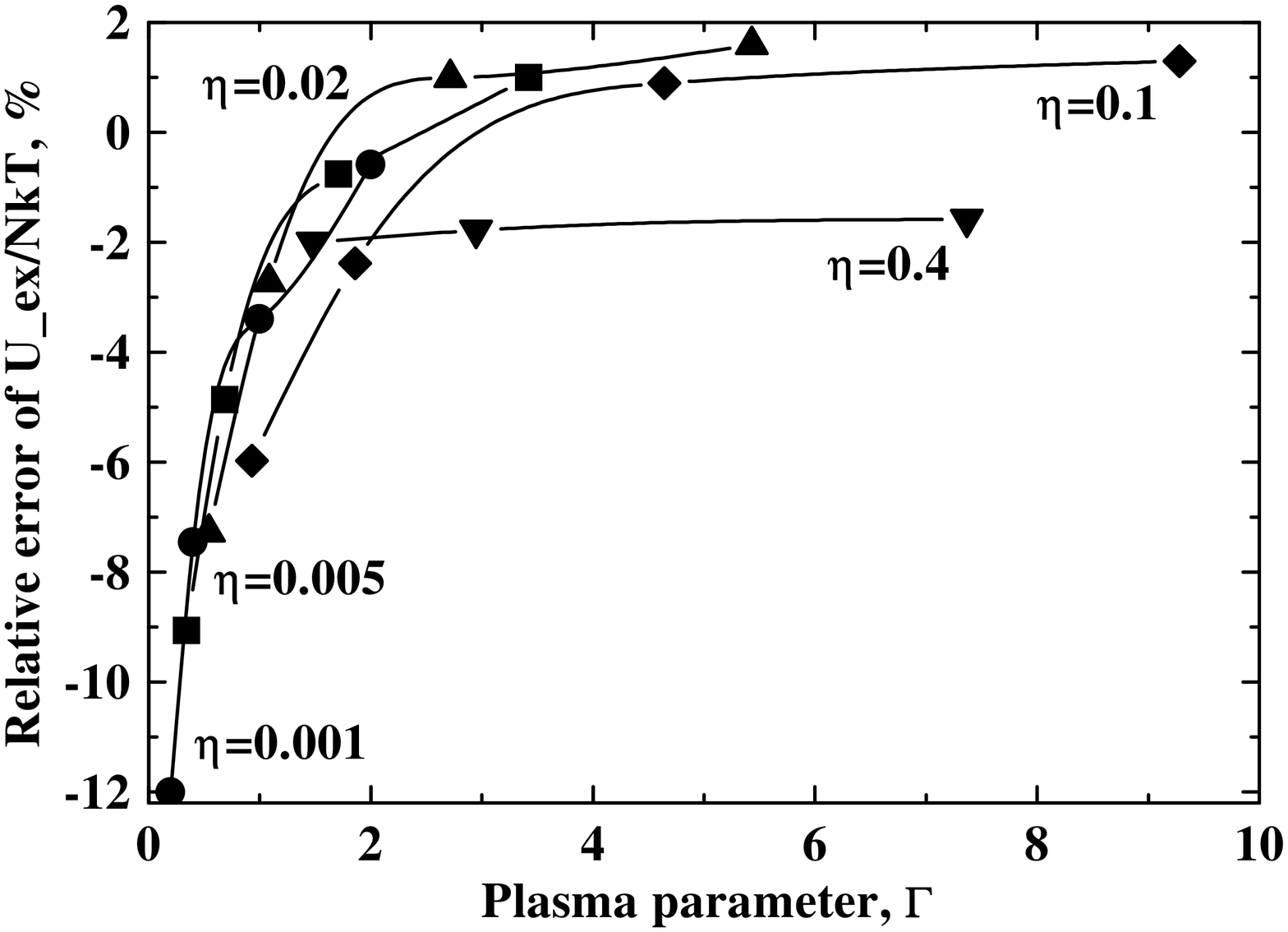}
\caption{Gives the relative error of the analytical expression
(\ref{U_fin}) for excess internal energy of the HCOCP
$U_{ex}/Nk_BT$ as a function of the plasma parameter $\Gamma$.
Notations are the same as in Fig.1. }\label{hc_err}
\end{figure}

\newpage

\begin{figure}[ht]
\centering
\includegraphics[scale=0.5,angle=270]{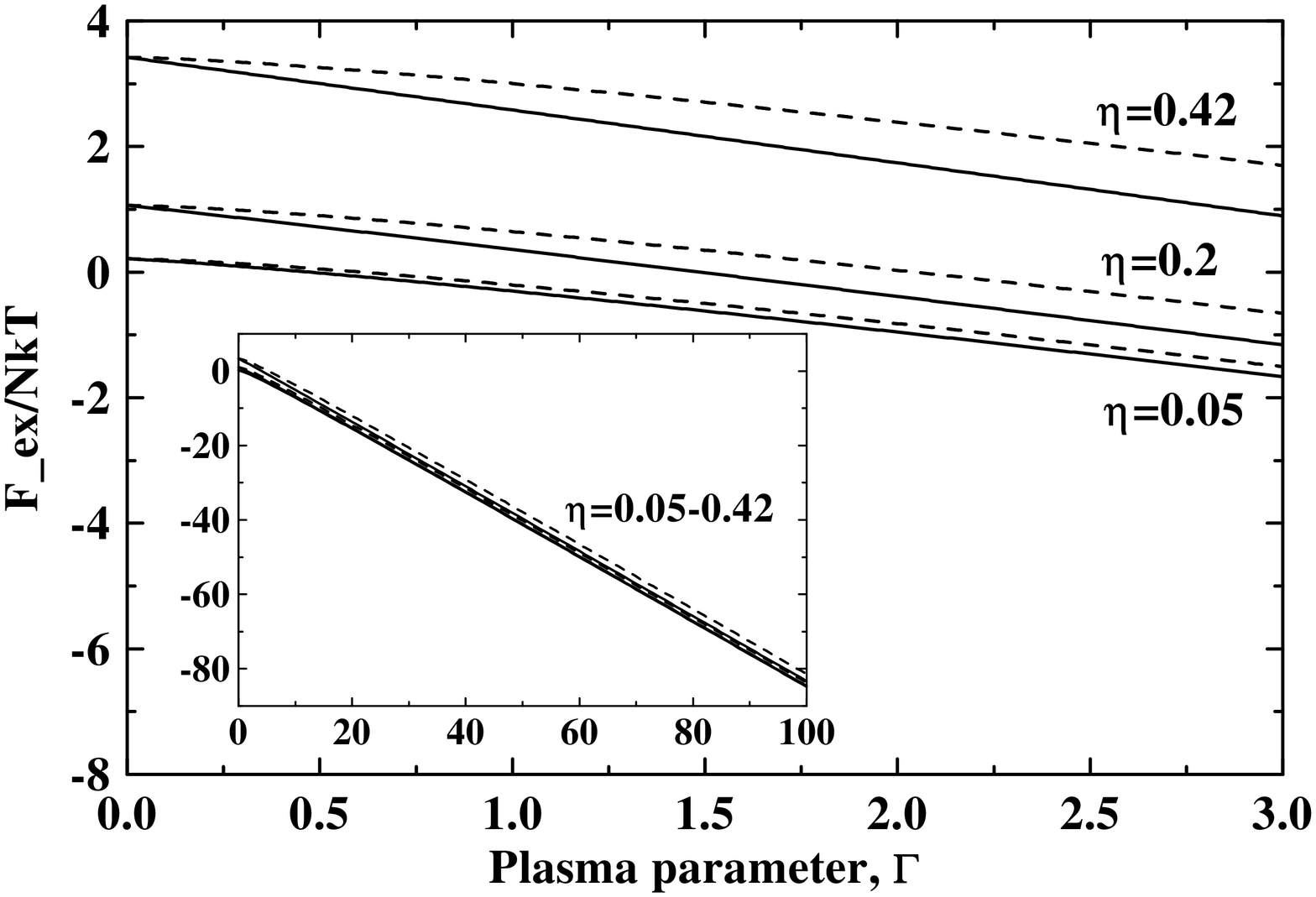}
\caption{Compares the ''complete'' equation of state,
$F_{ex,HCOCP}$ (\ref{Fex_fin}), for the free energy of the HCOCP
(solid lines) with the approximate one, given as the sum of the
free energy of the point particle OCP and that of the hard-core
fluid, $F_{ex,OCP}+F_{hc}$ (dashed lines). The curves from top to
bottom correspond respectively to $\eta=0.42$, $\eta=0.2$ and
$\eta=0.05$. In the inset the same dependence is shown for larger
range of $\Gamma$.}\label{nonad3}
\end{figure}


\begin{thebibliography}{99}
\bibitem{explain1}
The term with ${\bf k}=0$ conceals with the terms describing the
self-interaction of the compensating background and interaction of
the background with particles.
\bibitem{explain4}
The direct correlation function for the hard sphere fluid reads
\cite{WT}: $c_2(r)=0$, when $r\ge d$ and
$
c_2(r)=\displaystyle (1-\eta)^{-4}\left\{-(1+2\eta)^2 +6\eta\left
(1+\frac{1}{2}\eta\right )^2 (r/d) -\frac{1}{2}\eta (1+2\eta)^2
(r/d)^{3}\right \}
$
when $r<d$, where $d$ is the diameter of the hard spheres. For
$\tilde c_2({\bf k})$, which may be easily found from $c_2(r)$ we
do not need the complete expression.
\bibitem{c0comm}
The value of $\tilde c_2(0)$ may be obtained directly as the
zero-order moment of the Wertheim-Thiele solution  for $c_2(r)$
\cite{explain4} of the hard-sphere fluid. This is however based on
the approximate Percus-Yevik closure \cite{Gubbins} and occurs to
be less accurate than that related to $\tilde h_2(0)$, with
$\tilde h_2(0)$ taken from the Carnahan-Starling compressibility.
This happens because the Carnahan-Starling semiempirical equation
of state \cite{CarnHS} is more accurate than those found from the
Wertheim-Thiele solution \cite{Gubbins}.
\bibitem{explain5}
Note that the restriction $\varphi_{-{\bf k}}=\varphi^{*}_{{\bf
k}}$ does not reduce the total number of {\it physically
different} modes, since the modes with the wave-vectors ${\bf k}$
and  $-{\bf k}$ correspond to that spreading in the opposite
directions.
\end{thebibliography}
\end{document}